\documentclass[12pt]{article}
\usepackage{epsfig}
\usepackage{citesort}
\usepackage{amssymb,amsfonts}
\newcommand{\beq}{\begin{equation}}
\newcommand{\eeq}{\end{equation}}
\newcommand{\bea}{\begin{eqnarray}}
\newcommand{\eea}{\end{eqnarray}}

\newcommand{\Lagr}{{\mathcal L}}
\newcommand{\dd}{\mathrm d}

\begin{document}
\title{{\bf The NJL model and strange quark matter}}
\author{
C. Ratti
\vspace*{0.3cm}\\
\begin{tabular}{c}
{\it Dipartimento di Fisica Teorica, Universit\`a di Torino}\\
{\it and INFN, Sezione di Torino, 
Via P. Giuria 1, 10125 Torino, Italy}
\\
\end{tabular}
}

\date{\today}
\maketitle
\begin{abstract}
The stability of strange quark matter is studied within the Nambu Jona-Lasinio
 model with three different parameter sets. The model Lagrangian contains  
4-fermion (with and without vector interaction) and 6-fermion terms; the 
minimum energy per baryon number as a function of the 
strangeness fraction of the system is compared to the masses of hyperons having
the same strangeness fraction, and coherently calculated in the same version 
of the model, and for the same parameter set. The results show that in none of
 the different parameter sets strangelets are stable, and in some cases a
minimum in the energy per baryon does not even exist.
\end{abstract}

\vspace*{1cm}
\newpage

Strangelet detection in heavy ion collisions has been proposed long ago 
\cite{chin79,liu84,greiner87,greiner88,greiner91} as a signature of Quark
Gluon Plasma (QGP) formation: it has been suggested that rather cold droplets 
of stable or metastable strange-quark matter may be distilled in heavy-ion 
collisions during the phase transition from a baryon-rich QGP to hadron 
matter: the prompt anti-kaon and pion emission from the surface of the 
fireball could rapidly cool the QGP, thus favouring the condensation into 
metastable or stable droplets of strange quark matter
~\cite{greiner88,greiner91}.  

The investigation of strangelet stability is therefore rather crucial: 
it is fundamental to find out whether this system is more or 
less stable than hyperons, in order to understand which state is more likely 
to be produced in heavy ion collisions, either hyperons and strange mesons or
 strangelets. 
The properties and stability of strangelets have been discussed
within the MIT bag model, a confined gas stabilized by the 
vacuum pressure $B$~\cite{fahri84,greiner88,gilson93,schaffner97,madsen00}; 
more recent calculations, employing different quark models
~\cite{mishustin01,alberico02,wang02}, pointed out that strangelet stability
is strongly model dependent.

In this paper we want to discuss the properties of strange quark matter within
the NJL model, which has already been successfully utilized in the past to 
study strange quark matter~\cite{buballa99}; we consider homogeneous quark 
matter made up of $u$, $d$ and $s$
quarks; no $\beta$-equilibrium is required: in the system there is a 
strangeness fraction $R_s=\rho_s/\rho$, $\rho$ being the total baryon density 
of quarks and $\rho_s$ the baryonic density of strange quarks. Furthermore, 
electromagnetic interaction has been neglected, so that the minimum of the 
energy corresponds to an equal number of $u$ and $d$ quarks. 
We therefore consider the curves corresponding to the minimum energy per 
baryon number as a function of $R_s$ and compare these curves to the hyperon
masses evaluated in the same version of the NJL model and for the same 
parameter values, in order to understand whether strange matter is more stable 
than hyperons or viceversa.

Many versions of the NJL model have been used in the past; in our calculation 
we use a three flavour Lagrangian of the form 
\cite{vogl91,klevansky92,hatsuda94,rehberg96}:
\beq
{\Lagr}_{NJL}={\Lagr}_0+{\Lagr}_m+{\Lagr}_{(4)}+{\Lagr}_{(6)}
\label{Lagreq}
\eeq
where:
\bea
\Lagr_0 &=& i \,\bar{\psi}\, \gamma^\mu\,\partial_\mu \psi\, ,
\label{L0}\\
\Lagr_m &=& - \,\bar{\psi}\,{\bf\hat{m}}\,\psi\,,
\label{Lm}\\
\Lagr_{(4)}&=& \frac{G}{2} \sum\limits_{k=0}^8 \,
\left[\left(\bar{\psi}\lambda_k\psi\right)^2+
\left(i\bar{\psi}\gamma_5\lambda_k\psi\right)^2\right]\,+\\
\nonumber
&&-\frac{G_V}
{2}\sum\limits_{k=0}^8\left[\left(\bar{\psi}\gamma_{\mu}(\lambda_k)\psi
\right)^2+
\left(\bar{\psi}\gamma_{\mu}\gamma_5(\lambda_k)\psi\right)^2\right],
\label{L4}\\
\Lagr_{(6)}&=& -K\left[\det_{i,j}\left(\bar{\psi}_i(1+\gamma_5)\psi_j\right)+
\det_{i,j}\left(\bar{\psi}_i(1-\gamma_5)\psi_j\right)\right]\,.
\label{L6}
\eea
In the above
$\psi \equiv \left( {{\begin{array}{c} u \\ d \\ s\end{array}}} \right)$ 
is the quark field,
${\bf\hat{m}}    \equiv\left({{\begin{array}{ccc}
m_u & 0 & 0 \\0&m_d&0\\0&0&m_s\end{array}}}\right)$ is the mass matrix,
$\lambda_1\ldots\lambda_8$ are the Gell--Mann flavour matrices,
and $\lambda_0\equiv\sqrt{\frac{2}{3}}{\mathcal I}$.

$\Lagr_{(4)}$ generates four--leg interaction vertices, while 
$\Lagr_{(6)}$ gives rise to six--leg, flavour-mixing, interaction vertices;
 $G$ and $G_V$ 
are parameters of the model with the dimensions of $[L^{2}]$ and $K$ is a 
parameter with the dimensions of $[L^{5}]$. In the following we will consider 
both zero and nonzero values for $G_V$, in order to see the effects of a 
repulsive vector interaction on the stability of strangelets.
In the limit $m_i=0,~\forall i$, the symmetries of the model Lagrangian are 
the following ones:
\beq
U_V(1)\times SU_V(3)\times SU_A(3)\times SU_c(3)
\eeq
where, of course, $SU_c(3)$ is global; $U_A(1)$ is broken by
the existence of the axial anomaly.

In the mean field approximation it is possible to evaluate the dynamical quark
masses, generated by their interaction with the vacuum, which are given by
the following gap equation:
\beq
m_i^* = m_i-2G\langle\bar{\psi}_i \psi_i\rangle+2 K 
\langle\bar{\psi}_j \psi_j\rangle\langle\bar{\psi}_k \psi_k\rangle
\qquad\quad (i\ne j\ne k)
\label{mstari}
\eeq
where
\beq
\langle\bar{\psi}_i \psi_i\rangle = -\frac{3}{\pi^2}
\int\limits_{p_F^i}^\Lambda
\dd p \frac{m_i^* p^2}{\sqrt{p^2+(m_i^*)^2}}
\label{condens}
\eeq
 is the chiral condensate for the $i$-flavour.
Since the NJL model is not renormalizable, a three--dimensional 
regularization with an ultraviolet cut--off $\Lambda$ is introduced.
 
The energy density of the system in mean field approximation turns out to be:

\bea
\varepsilon=-&\sum\limits_{i=u,d,s}&\frac{3}{\pi^2}\int\limits_{p_F^i}^\Lambda
\dd p p^2\sqrt{p^2+(m_i^*)^2} +
\left[\sum\limits_{i=u,d,s} \,G
\langle\bar{\psi}_i\psi_i\rangle^2\right]+\\
\nonumber
&\sum\limits_{i=u,d,s}&G_V\rho_{V_i}^2
-4K\langle\bar{
\psi}_u\psi_u\rangle \langle\bar{\psi}_d\psi_d\rangle \langle\bar{\psi}_s
\psi_s\rangle+
\varepsilon_0.
\label{enden}
\eea
In the above formula, the constant $\varepsilon_0$ is introduced in order to 
set the vacuum energy density equal to zero, and $\rho_{V_i}$ is the time 
component of the vector current: 
$\rho_{V_i}=\langle\bar{\psi}_i\gamma_0\psi_i\rangle$.
The dependence of the above formula on $R_s$ and $\rho$ can be easily found 
by recalling that:
\bea
\rho_s&=&R_s\rho\\
\rho_{V_u}&=&\rho_{V_d}=3\frac{\left(1-R_s\right)}{2}\rho\\
\rho_{V_s}&=&3R_s\rho
\eea
in the above formulas, $\rho_i$ are the baryonic densities of quark $i$, which
are related to $\rho_{V_i}$ by:
\beq
\rho_i=\frac13\rho_{V_i}
\eeq
and to the Fermi momenta by:
\bea
k_{F_s}&=&\left(3\pi^2\rho_s\right)^{1/3}\\
k_{F_{u,d}}&=&\left(\frac{3\pi^2}{2}\rho\left(1-R_s\right)\right)^{1/3},
\eea
$\rho$ being the total baryon number density in the system ($\rho=N/V$).
In the above the colour degeneracy and baryon number $1/3$ of the quarks 
have been taken into account\footnote{Colour degeneracy is taken into account, 
even if no explicit gluon field is present in the model Lagrangian.}
.

From the above formulas, the energy per baryon number turns out to be:
\beq
\frac{E_{tot}}{N}=\frac{\epsilon_{tot}}{\rho}.
\label{Eperpart}
\eeq

In order to investigate the influence of the parameter values on strangelet 
stability, in this work three different parameter sets will be used:

\begin{center}
\begin{tabular}{|c|c|c|}
\hline
{\bf set 1} &{\bf set 2} &{\bf set 3}\\
\hline
$m\equiv m_u=m_d=5.5$ MeV & $m\equiv m_u=m_d=5.5$ MeV & $m\equiv m_u=m_d=3.6$ 
MeV\\ 
$m_s=140.7$ MeV           & $m_s=135.7$ MeV & $m_s=87$ MeV\\
$\Lambda=602.3$ MeV          & $\Lambda=631.4$ MeV & $\Lambda=750$ MeV\\
$G\Lambda^2=3.67$        & $G\Lambda^2=3.67$ & $G\Lambda^2=3.67$\\
$K\Lambda^5=12.36$        & $K\Lambda^5=9.29$ & $K\Lambda^5=8.54$\\
\hline
\end{tabular}
\end{center}
These parameters have been employed in Refs.~\cite{rehberg96} (set1), 
\cite{hatsuda94} (set 2) and \cite{klimt90} (set 3); in the first two cases, 
the current masses for the 
$u$ and $d$ quarks 
are fixed on the basis of isospin symmetry and of limits on the average
$\bar m=(m_u+m_d)/2$ at 1~GeV scale, while the remaining four parameters
are fitted to reproduce the masses of $\pi$,
$K$ and $\eta'$ mesons, together with the pion--decay constant 
$f_\pi$; the third parameter set also reproduces fairly well these 
experimental values.\\  
We choose for $G_V$ the two following values, in order to discuss the 
relevance of the repulsive vector 
interaction in strangelet formation: 

\beq
G_V=0~~~~~~~~~~~~~~~~~~~~~~~~~~~~G_V=0.5G
\eeq	
the second value being motivated, for example, in Ref.~\cite{vogl91}.

With the above parameter values, the effective quark masses in the vacuum and 
the chiral condensates turn out to be:

\begin{center}
\begin{tabular}{|c|c|c|c|}
\hline
&{\bf set 1}(MeV) &{\bf set 2}(MeV) &{\bf set 3}(MeV)\\
\hline
$m^*_u=m^*_d$ & 367.7 & 335.5 & 361\\
$m^*_s$ & 549.5 & 527.6 &501\\
$|\langle{\bar \psi}\psi\rangle_u|^{1/3}=|\langle{\bar \psi}\psi\rangle_d|
^{1/3}$ & 241.9 
& 246.7 & 287\\
$|\langle{\bar \psi}\psi\rangle_s|^{1/3}$ & 257.7 & 266.7 & 306\\
\hline
\end{tabular}
\end{center}

\begin{figure}
\begin{center}
\mbox{\epsfig{file=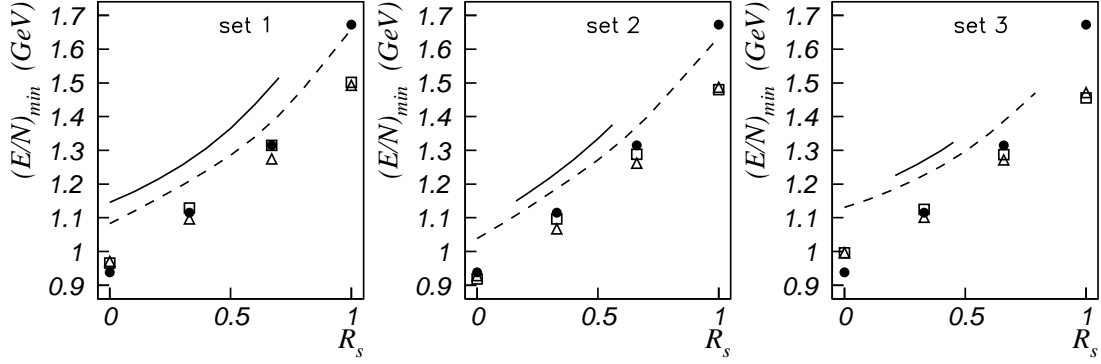,width=\textwidth}} 
%\textheight}}
\end{center}
\vskip -0.8cm
\caption{Minimum energy per baryon number as a function of $R_s$ for the three
different parameter sets, for $G_V=0$ (dashed lines) and $G_V=0.5G$ (continuous
 lines). Full circles are the experimental hyperon masses, the other dots are 
the theoretical hyperon masses corresponding to $G_V=0$ (empty triangles) and 
$G_V=0.5G$ (empty squares) respectively.}
\label{fig1}
\end{figure}

\begin{figure}
\begin{center}
\mbox{\epsfig{file=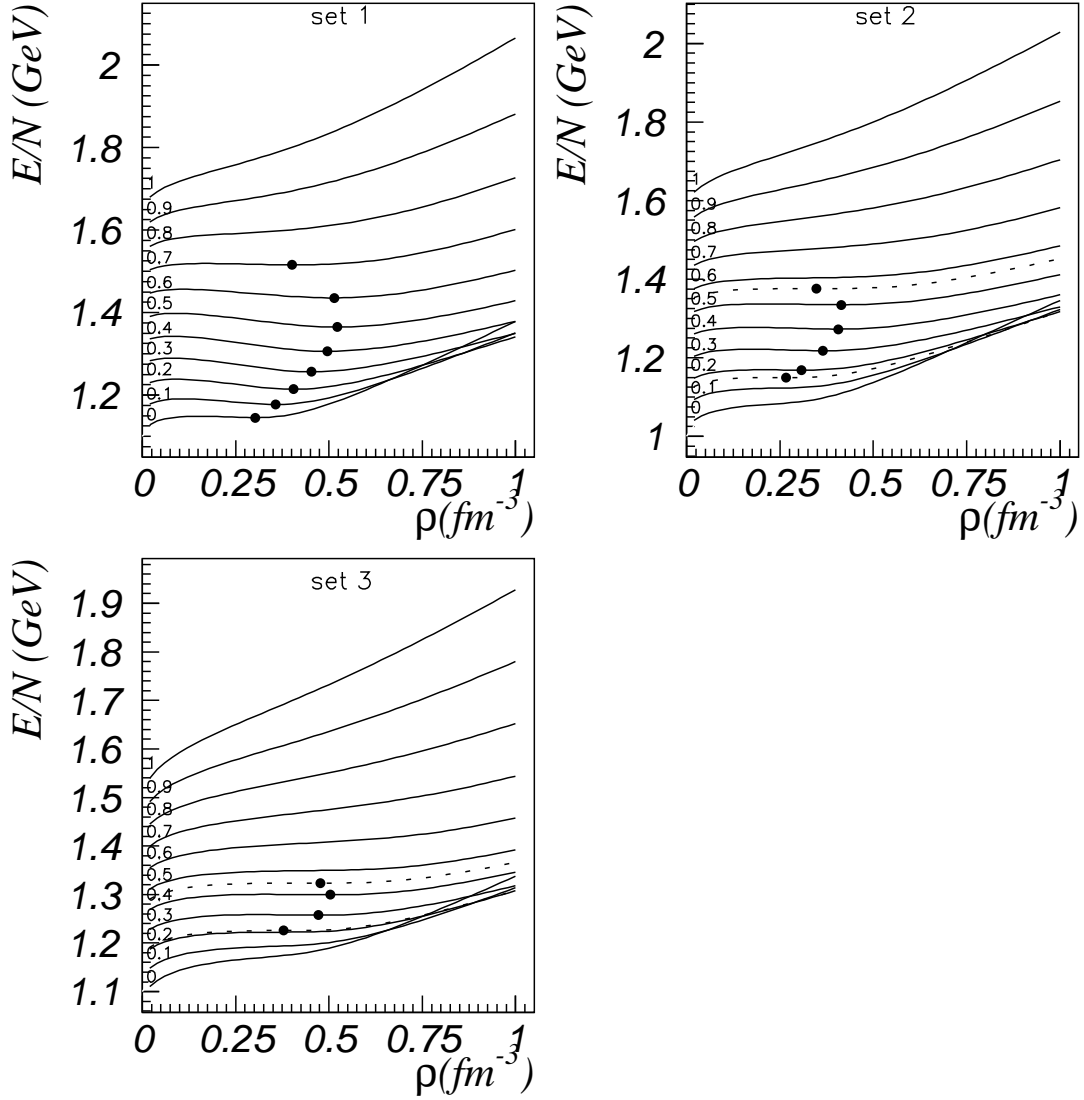,width=\textwidth}} 
%\textheight}}
\end{center}
\vskip -0.8cm
\caption{Energy per baryon number as a function of $\rho$ for different values
of $R_s$: the three panels correspond to the three
different parameter sets and $G_V=0.5G$. Full circles indicate local minima.
The dashed lines in the second and third panel indicate the first and last
 values of $R_s$ corresponding to a minimum: $0.16\leq R_s\leq 0.56$, and 
$0.21\leq R_s\leq 0.45$, respectively.}
\label{fig2}
\end{figure}

In Ref.~\cite{alberico02} a comparison is made between the minimum energy per 
baryon with respect to $\rho$ at fixed $R_s$, and the hyperon masses 
coherently calculated in the same model and for the same parameter values, in 
order to understand which state is more stable.
In the same spirit of this work, we compare our curves to the 
theoretical hyperon masses evaluated in the NJL model and for the same 
parameter values: for the first two parameter sets and $G_V=0$ we used the  
hyperon masses evaluated in Ref.~\cite{giacosa02}; following the same 
techniques presented in Ref.~\cite{giacosa02}, we also calculated the hyperon 
masses for parameter set 3 and for $G_V=0.5G$: they are all shown in Table~
\ref{tab1}.
In Fig.~\ref{fig1} we present our results for the three different parameter 
sets: the continuous lines correspond to $G_V=0.5G$, the 
dashed lines to $G_V=0$; concerning the hyperon masses, empty triangles 
correspond
 to $G_V=0$ and empty squares to $G_V=0.5G$; full circles are instead the 
experimental hyperon masses.
As it is evident from this figure, for all three parameter sets, with or 
without vector interactions, the curves corresponding to strangelets turn out 
to be well above the hyperon masses having the same strangeness fraction: 
independently of the parameter set used, and of the presence of the vector
interaction, strangelets are, therefore, not stable in the NJL model. 
From the left panel we can see that with the parameter set one and a nonzero 
value for $G_V$, a minimum in the energy per baryon is present 
 only up to $R_s=0.7$, as already pointed out in Ref.~\cite{mishustin01}.
In the central panel the results corresponding to parameter set 2 are shown, 
from which it is evident that the vector interaction further reduces the 
range of values of $R_s$ corresponding to a minimum in the energy per baryon:
in this case in fact a minimum is present only for 
$0.16\leq R_s\leq 0.56$.
Within parameter set 3 this range is even more reduced, as it appears from the
third panel: in the case of $G_V\neq 0$ a minimum is present only for 
$0.21\leq R_s\leq 0.45$. With this parameter set, even for $G_V=0$, the range 
of $R_s$ which is compatible with the existence of a minimum is limited 
to: $R_s\leq0.79$.
In order to get a feeling of the occurrence of these minima, we show the 
curves corresponding to the energy per 
baryon as a function of $\rho$ for different values of $R_s$ in Fig~\ref{fig2},
for the three parameter sets and with $G_V=0.5G$: full circles indicate local
minima.

Our analysis shows that the existence of stable or metastable strangelets is
 not 
supported by the NJL model: the curves corresponding to the
minimum energy per baryon as a function of the strangeness fraction are always
higher than the corresponding hyperon masses coherently calculated in the model
for the same parameter values, and in some cases the minimum does not exist;
these results seem therefore to indicate that hyperons are more likely to be
produced in heavy ion collisions, since they are more stable. This fact 
confirms the model dependence of the strangelet stability, which could set 
serious challenges to the search for these objects in heavy ion collisions.


\begin{table}[b]
\begin{center}
\begin{tabular}{cc|cccc}
\hline
\hline
 & Baryon & N &$\Lambda$ &$\Xi^{0}$ &$\Omega^{-}$ \\
\hline 
 &$m_{exp}$~(MeV) & 938.27 &1115.68 &1314.9 &1672.45 \\
\hline 
 &$m_{set\, 1}$~(MeV) & 970.86 &1096.34 &1274.51 &1493.10 \\
$G_V=0$ &$m_{set\, 2}$~(MeV) & 928.03 &1067.12 &1261.68 &1486.26\\
 &$m_{set\, 3}$~(MeV) & 996.77 &1101.34 &1271.78 &1471.33\\

\hline
 &$m_{set\, 1}$~(MeV) & 965.78 &1128.53 &1314.8 &1501.91 \\
$G_V=0.5G$ &$m_{set\, 2}$~(MeV) & 918.78 &1096.21 &1287.73 &1479.47\\
 &$m_{set\, 3}$~(MeV) & 996.65 &1124.97 &1287.6 &1455.77\\
\hline
\hline
\end{tabular}
\end{center}
\footnotesize
\caption{Experimental masses and theoretical masses of
hyperons calculated in the NJL model.}
\label{tab1}
\end{table}


\begin{thebibliography}{99}

\bibitem{chin79}
S.A. Chin and A.K. Kerman, Phys. Rev. Lett. {\bf 43} (1979) 1292.
%
\bibitem{liu84}
H. Liu and G.L. Shaw, Phys. Rev. {\bf D30} (1984) 1137.
%
\bibitem{greiner87}
C. Greiner {\it et al.}, Phys. Rev. Lett. {\bf 58} (1987) 1825.
%
\bibitem{greiner88}
C. Greiner {\it et al.},
Phys. Rev. {\bf D 38} (1988) 2797.
%
\bibitem{greiner91}
C. Greiner and H. St\"ocker, Phys. Rev. {\bf D44} (1991) 3517.
%nucl-th/9801062.
\bibitem{fahri84}
E. Fahri and R.L. Jaffe, Phys. Rev. {\bf D30} (1984) 2379.
%
\bibitem{gilson93}
E.P. Gilson and R.L. Jaffe, Phys. Rev. Lett. {\bf 71} (1993) 332.
%
\bibitem{schaffner97}
J.~Schaffner-Bielich {\it et al.},
%``Detectability of strange matter in heavy ion experiments,''
Phys.\ Rev.{\bf C 55} (1997) 3038.
%
\bibitem{madsen00}
J. Madsen, Phys. Rev. Lett. {\bf 85} (2000) 4687.
%
\bibitem{mishustin01}
I.N. Mishustin {\it et al.}, Phys. Atom. Nucl. 
{\bf 64} (2001) 802.
% 
\bibitem{alberico02}
W.M. Alberico {\it et al.}, Nucl. Phys. {\bf A706} (2002) 143. 
%
\bibitem{wang02}
P. Wang {\it et al.}, hep-ph/0205251.
%
\bibitem{buballa99}
M. Buballa and M. Oertel, Phys. Lett. {\bf B457} (1999) 261.

\bibitem{vogl91}
U. Vogl and W. Weise, Prog. Part. Nucl. Phys. {\bf 27} (1991) 195.
%
\bibitem{klevansky92}
S.P. Klevansky, Rev. Mod. Phys. {\bf 64} (1992) 649.
%
\bibitem{hatsuda94}
T. Hatsuda and T. Kunihiro, Phys. Rept. {\bf 247} (1994) 221.
%
\bibitem{rehberg96}
P.Rehberg {\it et al.}, Phys. Rev. {\bf C 53} (1996) 410.
%
\bibitem{klimt90}
S. Klimt {\it et al.} Phys. Lett. {\bf B 249} (1990) 386.
%
\bibitem{giacosa02}
W.M. Alberico {\it et al.}, nucl-th/0206071.
\end{thebibliography}
\end{document}